\documentclass[usenatbib]{mn2e} \usepackage{psfig}

\title[ULTRACAM photometry of OU Vir] {ULTRACAM photometry of the
  eclipsing cataclysmic variable OU Vir}

\author[W.\,J.\,Feline et al.] {W.\,J.\ Feline,$^1$\thanks{E-mail:
  w.feline@shef.ac.uk} V.\,S.\ Dhillon,$^1$ T.\,R.\ Marsh,$^{2,3}$ M.\,J.\
  Stevenson,$^1$ C.\,A.\ Watson$^1$ \newauthor and C.\,S.\ Brinkworth$^{3}$
  \\ $^1$Department of Physics and
  Astronomy, University of Sheffield, Sheffield, S3 7RH, UK\\
  $^2$Department of Physics, University of Warwick, Coventry CV4 7AL, UK\\
  $^3$Department of Physics and Astronomy, University of Southampton,
  Southampton, SO17 1BJ, UK\\}

\date{\center{\Large Accepted for publication in the Monthly
    Notices of the Royal Astronomical Society}}

\begin{document}
  \maketitle
  
\begin{abstract}
We present high-speed, three-colour photometry of the faint eclipsing
cataclysmic variable OU Vir. For the first time in OU Vir, separate
eclipses of the white dwarf and bright spot have been observed. We use
timings of these eclipses to derive a purely photometric model of the
system, obtaining a mass ratio of $q=0.175 \pm 0.025$, an inclination
of $i=79\fdg2 \pm 0\fdg7$ and a disc radius of $R_{d}/a=0.2315
\pm 0.0150$. We separate the white dwarf eclipse from the lightcurve
and, by fitting a blackbody spectrum to its flux in each passband,
obtain a white dwarf temperature of $T=13900 \pm 600$ K and a distance
of $D=51 \pm 17$ pc. Assuming that the primary obeys the
\citet{nauenberg72} mass-radius relation for white dwarfs and allowing
for temperature effects, we also find a primary mass
$M_{w}/M_{\sun}=0.89 \pm 0.20$, primary radius $R_{w}/R_{\sun}=0.0097
\pm 0.0031$ and orbital separation $a/R_{\sun}=0.74 \pm 0.05$.
\end{abstract}


\section{Introduction}
\label{introduction}

Cataclysmic variable stars (CVs) are short-period binary systems which
typically consist of a cool main-sequence star transferring mass via a
gas stream and accretion disc to a white dwarf primary. The impact
of the stream with the accretion disc forms a so-called `bright spot',
which in systems that are significantly inclined to our line of sight
can cause a rise in the observed flux as this region rotates into
view, resulting in an `orbital hump' in the lightcurve. In
high-inclination systems eclipses of the white dwarf, bright spot and
disc by the red dwarf secondary can also occur. Analysis of these
eclipses can yield determinations of system parameters such as the
mass ratio {\em q}, the orbital inclination {\em i} and the radius of
the accretion disc $R_{d}$ (e.g.\ \citealt{wood89a}). Eclipsing
systems are therefore valuable sources of data on CVs.

Dwarf novae are a sub-type of CVs which show intermittent luminosity
increases of 2--5 magnitudes, known as outbursts. A further sub-type of
dwarf novae are the SU UMa stars, which exhibit superoutbursts at regular
intervals, during which the luminosity increases by $\sim0.7$
magnitudes over the normal outburst maximum. These superoutbursts are
characterised by the presence of superhumps -- increases in brightness
that usually recur at a slightly longer period than the orbital cycle. There
is found to be a relationship between this superhump period excess
$\epsilon$ and the mass ratio \citep{patterson98}. Determinations of
the mass ratios of SU UMa stars are therefore useful to calibrate this
relation, which can then be used to determine the mass ratios of other
SU UMa stars.

OU Vir is a faint (V $\sim18$; \citealt{mason02}) eclipsing CV with a
 period of 1.75 hr which has been seen in outburst and probably
 superoutburst \citep{vanmunster00}, marking it as a SU UMa dwarf
 nova. \citet{mason02} presented time-resolved, multi-colour
 photometry and spectroscopy of OU Vir, concluding that the eclipse is
 of the bright spot and disc, but not the white dwarf. In this paper
 we present lightcurves of OU Vir, obtained with ULTRACAM, an
 ultra-fast, triple-beam CCD camera; for more details see
 \citet{dhillon01b}; Dhillon et~al., in preparation.


\begin{figure}
\centerline{\psfig{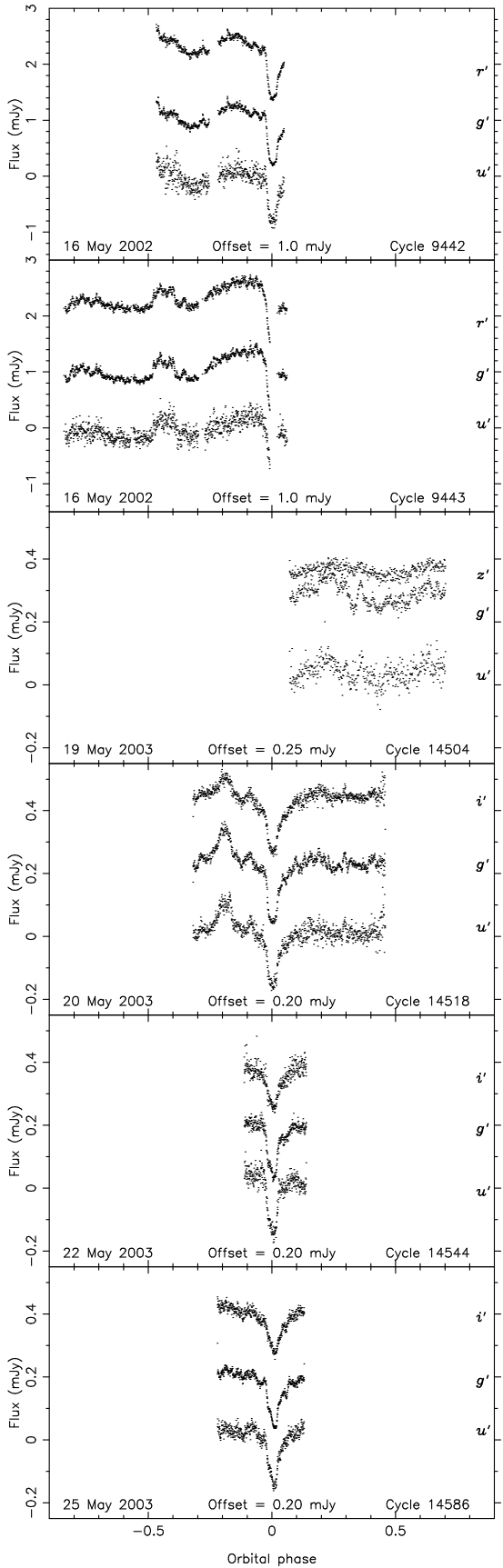}}
\caption{The lightcurve of OU Vir. The {\em r}$^{\prime}$, {\em
  z}$^{\prime}$ or {\em i}$^{\prime}$ data are offset vertically upwards
  and the {\em u}$^{\prime}$ data are offset vertically downwards.}
\label{lightcurve}
\end{figure}

\section{Observations}
\label{observations}

OU Vir was observed on the nights of 16 May 2002 and 19, 20, 22
and 25 May 2003 using  ULTRACAM  on the 4.2-m William Herschel
Telescope (WHT) at the Isaac Newton Group of Telecopes, La Palma. The
observations were obtained simultaneously in the {\em r}$^{\prime}$,
{\em g}$^{\prime}$, and {\em u}$^{\prime}$ colour bands on the 16 May
2002,  the {\em z}$^{\prime}$, {\em g}$^{\prime}$, and {\em
u}$^{\prime}$ bands on the 19 May 2003, and the {\em i}$^{\prime}$,
{\em g}$^{\prime}$, and {\em u}$^{\prime}$ bands on subsequent
nights. Data reduction was carried out using the ULTRACAM pipeline
data reduction software. Unfortunately, we were unable to observe a
standard star in the {\em z}$^{\prime}$ band using ULTRACAM, so the
zeropoint for this filter was chosen to be the same as for the {\em
i}$^{\prime}$ band; this will only affect the scale of the {\em
z}$^{\prime}$ lightcurve, not its shape. All the data were corrected
to zero airmass using the nightly extinction coefficients measured by
the Carlsberg Meridian Telescope on La Palma in the {\em r}$^{\prime}$ filter,
and converted to other colour bands using the procedure described by
\citet{king85} and the effective wavelengths of the filters given by
\citet{fukugita96}.

The lightcurves of OU Vir are shown in Figure \ref{lightcurve}. On
16 May 2002 the exposure time was 0.5 seconds and the data points were
separated by approximately 4.9 seconds. These data were obtained on
the first night of commissioning and hence were adversely affected by
typical commissioning problems, chiefly excess noise in the {\em
u}$^{\prime}$ band and limited time resolution due to the dead-time
between exposures. The data taken in 2003 had no such problems, and
hence the dead-time was reduced to 0.025 seconds. The exposure times
for these observations varied according to the seeing on the night,
and were 9.2 seconds on 19 May 2003, 5.2 seconds on 20 May 2003 and 4.2
seconds thereafter.

\begin{figure}
\centerline{\psfig{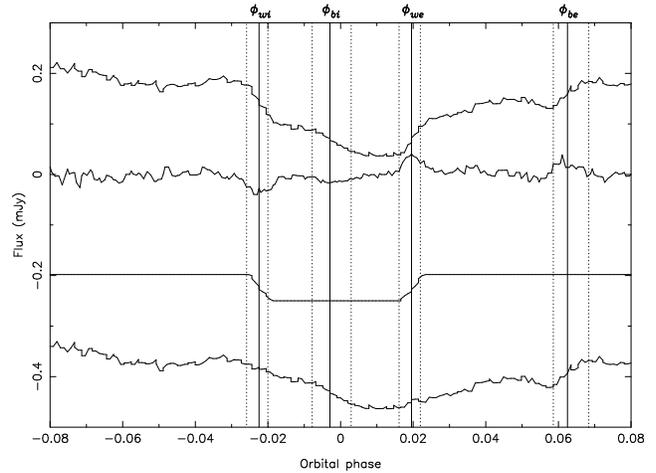}}
\caption{White dwarf deconvolution of the {\em g}$^{\prime}$ band
  lightcurve of 25 May 2003. Top to bottom: The data after smoothing
  by a median filter; the derivative after smoothing by a box car
  filter and subtraction of the spline fit to this, multiplied by a
  factor of 5 for clarity; the reconstructed white dwarf lightcurve,
  shifted downwards by 0.25 mJy; the original lightcurve minus the
  white dwarf lightcurve after smoothing by a median filter, shifted
  downwards by 0.5 mJy. The vertical lines show the contact phases of
  the white dwarf and bright spot eclipses, the dotted lines
  corresponding to $\phi_{w1}\ldots\phi_{w4}$,
  $\phi_{b1}\ldots\phi_{b4}$ and the solid lines (labelled) to
  $\phi_{wi}$, $\phi_{we}$ and $\phi_{bi}$, $\phi_{be}$. The bright
  spot ingress and egress are plainly visible, quickly following the
  white dwarf ingress and egress respectively.}
\label{deconvolution}
\end{figure}

\section{Lightcurve morphology}
\label{morphology}

OU Vir was observed during descent from superoutburst in both 2002 and
2003. \citet{mason02} found that for OU Vir out of eclipse and during
quiescence, $V=18.08$ and $B-V=0.14$, which corresponds to {\em
  g}$^{\prime}\sim0.2$ mJy \citep{smith02}. \citet{vanmunster00} quote
an outburst amplitude of approximately 4 magnitudes (corresponding to
a peak {\em g}$^{\prime}$ flux of $\sim8.4$ mJy). The system appears
to be significantly below superoutburst maximum on the 16 May 2002,
and almost at its quiescent brightness during the 2003 observations:
The maximum superoutburst flux we have observed is $\sim1.2$ mJy on 16
May 2002, and the out-of-eclipse quiescent flux is $\sim0.2$ mJy on 22
and 25 May 2003.

As the descent from superoutburst is typically much slower than the
ascent, we suspect the 2002 observations took place during the former
state, but of course we cannot be certain. The 2002 data are, as far
as we are aware, the only observations of this superoutburst. The 2003
superoutburst was first reported on 2 May \citep{kato03}. The
superhump is visible in both the 16 May 2002 lightcurve at phase
$\sim-0.45$ and the 20 May 2003 lightcurve at phase
$\sim-0.2$, but is not visible in later data, either because it
had faded or it was at a phase which was not observed.

The lower three panels of Figure \ref{lightcurve} show changes in the
eclipse morphology that occur during the descent from
superoutburst. Importantly, the eclipse morphology changed drastically
from a single sharp ingress and egress on  20 May 2003, to separate
eclipses of the white dwarf and bright spot in the {\em u}$^{\prime}$
and {\em g}$^{\prime}$ bands on 25 May 2003. The bright spot
ingress is also visible in the {\em g}$^{\prime}$ data of 22 May
2003. (The second egress feature visible in this lightcurve is more
gradual than it appears in the compressed scale of Figure
\ref{lightcurve}, and is mainly due to the egress of the disc.) This
evolution of the eclipse morphology is typical for a system going from
(super) outburst to quiescence (see, for example, \citealt{rutten92c}).

\begin{figure}
\centerline{\psfig{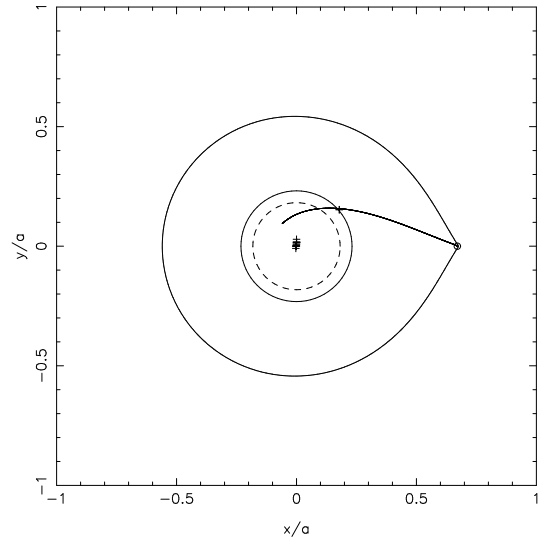}}
\caption{Trajectory of the gas stream from the secondary star for
  $q=0.175$ and $i=79\fdg2$. Top: The system with the primary Roche lobe,
  $L_{1}$ point and disc of radius $R_{d}=0.2315a$ plotted. The
  positions of the white dwarf and bright spot light centres
  corresponding to the observed ingress and egress phases for $q$ and
  $i$ as above are also plotted. The circularisation radius
  \citep[][ equation 13]{verbunt88} of $R_{circ}=0.1820a$ is shown as
  a dashed circle. The stream passes through the bright spot points
  (note that the timings of 22 May 2003 are of the ingress only
  which prevents it from being plotted).}
\label{mass}
\end{figure}

\begin{table*}
\begin{center}
\caption[]{Mid-eclipse timings.}
\begin{tabular}{ccccc}
\hline
Date & \multicolumn{4}{c}{HJD}\\
 & {\em u}$^{\prime}$ & {\em g}$^{\prime}$ & {\em r}$^{\prime}$ &
 {\em i}$^{\prime}$\\
\hline
16 May 2002 & 2452411.523977 & 2452411.523921 & 2452411.523892 & --\\
20 May 2003 & 2452780.580157 & 2452780.580278 & -- & 2452780.580217\\
22 May 2003 & 2452782.470534 & 2452782.470558 & -- & 2452782.470509\\
25 May 2003 & 2452785.524083 & 2452785.524083 & -- & 2452785.524255\\
\hline
\end{tabular}
\label{eclipse_times}
\end{center}
\end{table*}

\begin{table*}
\begin{center}
\caption[]{White dwarf contact phases and flux.}
\begin{tabular}{ccccccccc}
\hline
Date & Band & $\phi_{w1}$ & $\phi_{w2}$ & $\phi_{w3}$ & $\phi_{w4}$ &
$\phi_{wi}$ & $\phi_{we}$ & Flux (mJy)\\
\hline
16 May 2002  & {\em u}$^{\prime}$ & -0.024414 & -0.017578 &
0.018555 & 0.025391 & -0.020508 & 0.022461 & --\\
      & {\em g}$^{\prime}$ & -0.027344 & -0.018555 &
0.016602 & 0.025391 & -0.022461 & 0.021484 & --\\
      & {\em r}$^{\prime}$ & -0.028320 & -0.013672 & 0.012695
& 0.027344 & -0.020508 & 0.020508 & --\\
20 May 2003 & {\em u}$^{\prime}$ & -0.023438 & -0.016602 & 0.018555
& 0.025491 & -0.019531 & 0.022461 & --\\
      & {\em g}$^{\prime}$ & -0.025391 & -0.017578 & 0.017578
& 0.024414 & -0.021484 & 0.021484 & --\\
      & {\em i}$^{\prime}$ & -0.023438 & -0.014648 & 0.016602
& 0.025391 & -0.018555 & 0.021484 & --\\
22 May 2003 & {\em u}$^{\prime}$ & -0.024414 & -0.016602 & 0.017578
& 0.025391 & -0.020508 & 0.021484 & --\\
      & {\em g}$^{\prime}$ & -0.026367 & -0.018555 & 0.016602
& 0.024414 & -0.022461 & 0.020508 & --\\
      & {\em i}$^{\prime}$ & -0.021484 & -0.018555 & 0.016602
& 0.020508 & -0.019531 & 0.018555 & --\\
25 May 2003 & {\em u}$^{\prime}$ & -0.025391 & -0.016602 & 0.014648
& 0.023438 & -0.020508 & 0.019531 & 0.0537\\
      & {\em g}$^{\prime}$ & -0.025391 & -0.019531 & 0.016602
& 0.022461 & -0.022461 & 0.019531 & 0.0519\\
      & {\em i}$^{\prime}$ & -0.022461 & -0.018555 & 0.017578 &
0.022461 & -0.020508 & 0.020508 & 0.0146\\
\hline
\end{tabular}
\label{wd_times}
\end{center}
\end{table*}

\begin{table*}
\begin{center}
\caption[]{Bright spot contact phases.}
\begin{tabular}{cccccccc}
\hline
Date & Band & $\phi_{b1}$ & $\phi_{b2}$ & $\phi_{b3}$ & $\phi_{b4}$ &
$\phi_{bi}$ & $\phi_{be}$\\
\hline
22 May 2003 & {\em g}$^{\prime}$ & -0.002930 & 0.002930 & -- & -- &
0.000977 & --\\
25 May 2003 & {\em u}$^{\prime}$ & -0.007813 & 0.002930 &
0.060547 & 0.065430 & -0.002930 & 0.062500\\
      & {\em g}$^{\prime}$ & -0.007813 & 0.002930 &
0.058594 & 0.068359 & -0.002930 & 0.062500\\
\hline
\end{tabular}
\label{bs_times}
\end{center}
\end{table*}

\section{Eclipse contact phases}
\label{contact}

The white dwarf eclipse contact phases given in Tables
\ref{eclipse_times} and \ref{wd_times} were determined using the
techniques described by \citet*{wood85}, \citet{wood86b} and
\citet{wood89a}. A median filter was used to smooth the data, the
derivative of which was then calculated numerically. A box-car filter
was  applied to this derivative, and simple searches were made to
locate the minimum and maximum values of the derivative corresponding
to the midpoints of ingress $\phi_{i}$ and egress $\phi_{e}$. (In
fact this method locates the steepest part of the ingress and egress,
but we would expect these to correspond to the midpoints unless the
light distribution is asymmetrical.) If a bright spot eclipse is also
present, care must be taken to ensure that at this stage the ingress
and egress of the white dwarf are not confused with those of the
bright spot. The eclipse contact phases corresponding to the start and
end of the ingress $\phi_{1}$, $\phi_{2}$ and the start and end of the
egress $\phi_{3}$, $\phi_{4}$ were determined by locating the points
where the derivative differs significantly from a spline fit to the
more slowly varying component.

Once the white dwarf eclipse contact phases have been found, the
white dwarf lightcurve can be reconstructed and subtracted from the
overall lightcurve as illustrated in Figure \ref{deconvolution}. The
out-of-eclipse white dwarf fluxes thus found are given in Table
\ref{wd_times}. The white dwarf flux can be used to determine its
temperature and distance; see section \ref{sys_parameters}. Once this
has been done the bright spot eclipse contact phases (given in
Table \ref{bs_times}) can be determined by a similar method
\citep{wood89a} and its lightcurve removed from that of the disc
eclipse. If successful, this process can be used to determine the
bright spot temperature. Unfortunately we were unsuccessful in our
attempts to do this, probably because flickering hindered accurate
determination of the bright spot flux and contact phases.

In the discussion that follows we use the suffixes `$w$' and `$b$' to
denote white dwarf and bright spot contact phases, respectively (e.g.\
$\phi_{wi}$ means the mid-ingress point of the white dwarf eclipse).

\section{Orbital ephemeris}
\label{ephemeris}

A linear least-squares fit to the times of mid-eclipse given in Table
\ref{eclipse_times} (calculated using the techniques described in
section \ref{contact} and taking the midpoint of the white dwarf
eclipse as the point of mid-eclipse) and those of \citet[][ private
communication]{vanmunster00} gives the following ephemeris:

\begin{displaymath}
\begin{array}{ccrcrl}
\\ HJD & = & 2451725.03283 & + & 0.072706113 & \cdot\:E.\\ & & 7 & \pm
 & 5 &
\end{array} 
\end{displaymath} Errors of $\pm 4\times 10^{-5}$ days were used for
 the ULTRACAM data, and errors of $\pm 7\times 10^{-4}$ days for the
 \citet[][ private communication]{vanmunster00} data. This ephemeris
 was used to phase all of our data.

\section{System Parameters}
\label{sys_parameters}

The derivation of the system parameters relies upon the fact that
there is a unique relationship between the mass ratio and orbital
inclination for a given eclipse phase width $\Delta\phi = \phi_{we} -
\phi_{wi}$:
\begin {enumerate}
\item At smaller orbital inclinations a larger secondary radius
$R_{r}$ is required in order to produce a given eclipse width.
\item The secondary radius is defined by the mass ratio because the
secondary fills its Roche lobe.
\item Therefore for a specific white dwarf eclipse width, the
inclination is known as a function of the mass ratio.
\end{enumerate}

The shape of the system does not depend on the orbital separation $a$;
this just determines the scale. The orbital separation is determined
by assuming a mass-radius relation for the primary (see later).

\begin{figure}
\centerline{\psfig{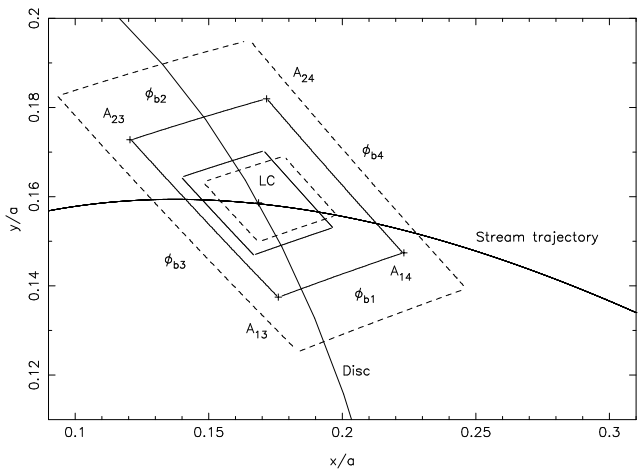}}
\caption{Horizontal structure of the bright spot for $q=0.175$,
  showing the region on the orbital plane within which the bright spot
  lies. The light centre {\em LC} is marked by a cross, surrounded by
  the inner solid box which corresponds to the rms variations in
  position. The phase arcs which correspond to the bright spot contact
  phases are shown as the outer solid box, with the rms variations in
  position shown as the two dashed boxes. As all the timings of
  $\phi_{b2}$ and $\phi_{be}$ are identical, the rms variations of
  $\phi_{b1}$ and $\phi_{bi}$ respectively have been used
  instead. Intersections of the phase arcs $\phi_{bj}$ and $\phi_{bk}$
  are marked $A_{\rmn{jk}}$, with crosses. The stream trajectory and
  disc of radius $R_{d}=0.2315a$ are also plotted as solid curves.}
\label{bs_horizontal}
\end{figure}

\begin{figure}
\centerline{\psfig{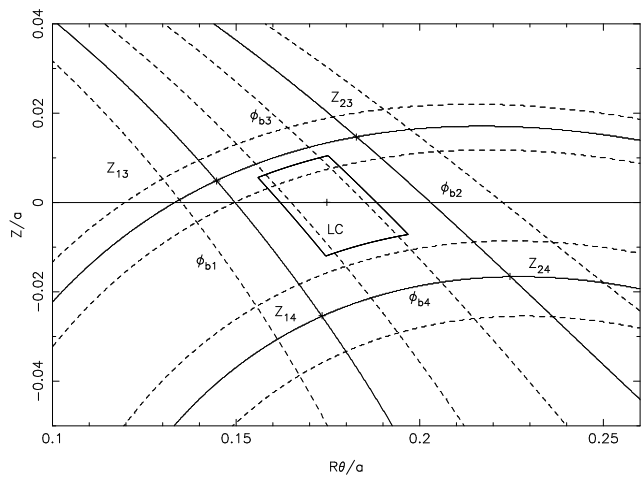}}
\caption{As Figure \ref{bs_horizontal}, but showing the vertical
  structure of the bright spot. The phase arcs are projected onto a
  vertical cylinder of radius $0.2315a$ (equal to that of the disc),
  i.e.\ the {\em x}--axis is stepping around the edge of the
  disc. $\theta$ is in radians. The intersections of the phase arcs
  $\phi_{bj}$ and $\phi_{bk}$ are marked $Z_{\rmn{jk}}$, with crosses.}
\label{bs_vertical}
\end{figure}

The trajectory of the gas stream originating from the inner Lagrangian
point $L_{1}$ is calculated by solving the equations of motion
\citep{flannery75} using a second-order Runge--Kutta technique and
conserving the Jacobi Energy to 1 part in $10^{4}$. This assumes that
the gas stream follows a ballistic path. Figure \ref{mass} shows a
theoretical gas stream for $q=0.175$. Figures \ref{bs_horizontal} and
\ref{bs_vertical} show expanded views of the bright spot region. As
$q$ decreases, the path of the stream moves away from the white
dwarf. For a given mass ratio $q$ each point on the stream has a
unique phase of ingress and egress.

For each phase, the limb of the secondary forms an arc when projected
along the line of sight onto a given plane (hereafter referred to as a
phase arc): each point on an individual phase arc is eclipsed at the same
time. The intersection of the phase arcs corresponding to the
respective eclipse contact phases can be used to constrain the size of
the white dwarf and the structure of the bright spot. The light
centres of the white dwarf and bright spot must lie at the
intersection of the phase arcs corresponding to the relevant phases of
mid-ingress and mid-egress, $\phi_{i}$ and $\phi_{e}$. The phase arcs
were calculated using full Roche lobe geometry rather than an
approximate calculation.

The mass ratio and hence the inclination may be determined by
comparing the bright spot light centres corresponding to the measured
eclipse contact phases $\phi_{wi}$ and $\phi_{we}$ with the
theoretical stream trajectories for different mass ratios $q$. This
requires the assumption that the gas stream passes directly through
the light centre of the bright spot. As illustrated in Figures
\ref{bs_horizontal} and \ref{bs_vertical}, we constrain the light
centre of the bright spot to be the point where the gas stream and
outer edge of the disc intersect, so that the distance from the
primary at which the gas stream passes through the light centre of the
bright spot gives the relative outer disc radius $R_{d}/a$.

The bright spot timings thus yield a mass ratio of $q=0.175 \pm 0.025$
and an inclination of $i=79\fdg2 \pm0 \fdg7$ for an eclipse phase
width $\Delta\phi = 0.041585$. The errors are determined by the rms
variations in the measured contact phases. Figures \ref{bs_horizontal}
and \ref{bs_vertical} show the eclipse constraints on the structure of
the bright spot. We use these to determine upper limits on the angular
size and the radial and vertical extent of the bright spot, defining
\begin{eqnarray}
\label{eq:theta}
\Delta \theta & = &
(\theta_{23}+\theta_{24}-\theta_{13}-\theta_{14})/2\\
\label{eq:r}
\Delta R_{d} & = & (R_{24}+R_{14}-R_{23}-R_{13})/2\\
\label{eq:z}
\Delta Z & = & (H_{23}-H_{14})/2,
\end{eqnarray}
where $R_{\rmn{jk}}$ and $\theta_{\rmn{jk}}$ are the radius and
azimuth of $A_{\rmn{jk}}$ and $H_{\rmn{jk}}$ the height of
$Z_{\rmn{jk}}$ above the orbital plane, as defined in Figures
\ref{bs_horizontal} and \ref{bs_vertical}. $\theta$ increases in the
direction of orbital motion and is zero at the line joining the
centres of the two stars. Note that the definition of $\Delta Z$ in
equation \ref{eq:z} differs slightly to that defined in
\citet{wood86b}: this is in order to be more consistent with the
definitions of $\Delta \theta$ and $\Delta R_{d}$ in equations
\ref{eq:theta} and \ref{eq:r}. The mean position and extent of the
bright spot are given in Table \ref{bs}.

From Figures \ref{bs_horizontal} and \ref{bs_vertical} we estimate
that the gas stream passing through the light centre of the bright
spot could just touch the phase arcs corresponding to $\phi_{b1}$ and
$\phi_{b4}$ for a stream of circular cross-section with a radius
$\varepsilon /a = 0.0175 \pm 0.0025$. The bright spot appears to be more
extended azimuthally than radially, which can be understood by the
shock front extending both up-disc and down-disc from the point of
impact. The size of the stream is similar to that expected from
theoretical studies \citep{lubow75, lubow76} and that obtained by
studies of similar objects \citep{wood86b,wood89a}, so our
assumption that the stream passes through the light centre of the
bright spot is reasonable.

\begin{figure}
\centerline{\psfig{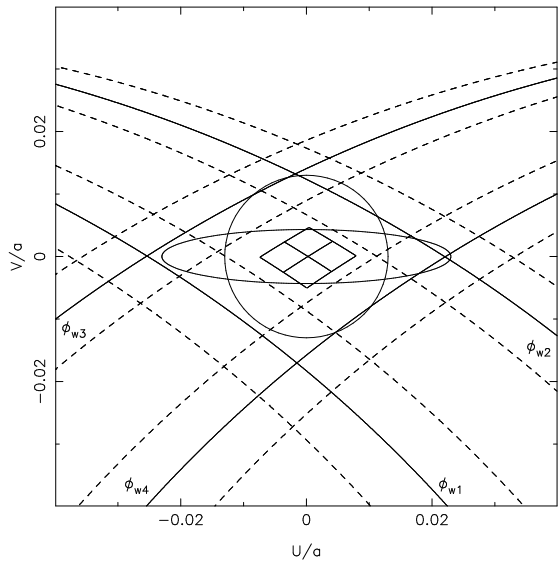}}
\caption{Projection of the white dwarf phase arcs onto the plane
  perpendicular to the line of sight. $U$ and $V$ are orthogonal
  coordinates perpendicular to the line of sight, $U$ being parallel
  to the binary plane. Solid curves correspond to the contact phases
  of the white dwarf, dotted curves to the rms variations of the phase
  arcs. The light centre is also plotted surrounded by the solid box
  corresponding to the rms variations in phase. The projection of the
  white dwarf and accretion belt centered on $U,V=0$ are shown for
  $R_{w}=0.013a$ and $R_{belt}=0.023a$.}
\label{whitedwarf}
\end{figure}

Figure \ref{whitedwarf} shows the eclipse constraints on the radius of
the white dwarf. Using the mass ratio and orbital inclination derived
earlier, we find that the white dwarf has a radius of $R_{w}=0.013 \pm
0.004a$. An alternative possibility is that the sharp eclipse is
caused by a bright inner disc region or boundary layer of radius
$R_{belt}=0.023 \pm 0.010a$ surrounding the white dwarf like a
belt. These errors are calculated using the rms variations in the
measured contact phases. Another possibility is that the lower
hemisphere of the white dwarf is obscured by an optically thick
accretion disc, which would result in the white dwarf radius being
$R_{w}\geq0.013a$. This is not the case, however, as the
contact phases $\phi_{wi}$ and $\phi_{we}$ lie half-way through the
white dwarf ingress and egress (see section \ref{contact}), i.e.\ the
light centre in Figure \ref{whitedwarf} lies at the origin and the
light distribution is symmetrical.

The following analysis assumes that the eclipse is of a bare white
dwarf. If the eclipse is actually of a belt and the white dwarf itself
is not visible, then the white dwarf radius must be smaller than
$R_{belt}=0.023a$. If the white dwarf does contribute significantly to
the eclipsed light, then we have the additional constraint that its
radius must be $R_{w} \leq 0.013a$, so that the white dwarf mass given
in Table \ref{parameters} is actually a lower limit. The only way to
verify the assumption that the central light source is the white
dwarf alone is to measure the semi-amplitude of the radial velocity
curve of the secondary star, $K_{r}$, and compare it to that predicted
by the photometric model in Table \ref{parameters}. We could also
check if this assumption is true using a longer baseline of quiescent
observations, as one might expect eclipse timings of an accretion belt
to be much more variable than those of a white dwarf. We note,
however, that the white dwarf mass given in Table \ref{parameters} is
consistent with the mean white dwarf mass of $0.69\pm0.13M_{\sun}$ for
CVs below the period gap \citep{smith98a}. Also, \citet*{baptista00}
point out that short-period dwarf novae (specifically OY Car, Z Cha
and HT Cas) like OU Vir tend to accrete directly onto the white dwarf,
whereas longer-period dwarf novae (IP Peg and EX Dra) usually have
boundary layers. We will continue under the assumption that the
central eclipsed object is indeed a bare white dwarf.

\begin{table}
\begin{center}
\caption[]{Mean position and extent of the bright spot as defined by
  equations \ref{eq:theta} -- \ref{eq:z}. $\Delta Z_{2}$ is calculated
  according to the definition used by \citet{wood86b}, for ease of
  comparison.}
\begin{tabular}{cc}
\hline
$\Delta R_{d}/a$ & 0.0417\\
$\Delta \theta$ & 15\fdg17\\
$\Delta Z/a$ & 0.0200\\
$\Delta Z_{2}/a$ & 0.0147\\
$R_{d}/a$ & 0.2315\\
$\theta$ & 43\fdg24\\
\hline
\end{tabular}
\label{bs}
\end{center}
\end{table}

To determine the remaining system parameters we have used the
Nauenberg mass-radius relation for a cold, non-rotating white dwarf
\citep{nauenberg72, cook84},
\begin{equation}
\label{eq:nauenberg}
R_{w}=7.795\times10^{6}\left[ \left(
  \frac{1.44M_{\sun}}{M_{w}} \right) ^{\frac{2}{3}} - \left(
  \frac{M_{w}}{1.44M_{\sun}} \right) ^{\frac{2}{3}} \right]
  ^{\frac{1}{2}}\rmn{m},
\end{equation}
which gives an analytical approximation to the Hamada--Salpeter
mass-radius relation \citep{hamada61}. If we set $R_{w}/a=y$, we can
rewrite Kepler's third law in terms of the parameters $R_{w}$
and $y$, giving another restriction on the white dwarf radius:
\begin{equation}
\label{eq:kepler}
R_{w}=y \left( \frac{GM_{w}(1+q)P^{2}}{4\pi^{2}} \right)
^{\frac{1}{3}}.
\end{equation}
Equations \ref{eq:nauenberg} and \ref{eq:kepler} can be easily solved
to give the system parameters, given in Table \ref{parameters}. The
secondary radius $R_{r}$ has been calculated by approximating it to
the effective radius of the Roche lobe \citep{eggleton83}:
\begin{eqnarray}
\label{eq:roche}
R_{\rmn{L}}=\frac{0.49aq^{\frac{2}{3}}}{0.6q^\frac{2}{3}+
  \rm{ln}\left(1+q^{\frac{1}{3}}\right)}\,,&&0<q<\infty
\end{eqnarray}
which is accurate to better than 1 per cent. As the
\citet{nauenberg72} mass-radius relation assumes a cold white dwarf,
we have attempted to correct this relation to a temperature of
$\sim14000$ K, the approximate temperature given by the
blackbody fit below. \citet{wood89a} and \citet{koester86}
note that the radius of a white dwarf at $10^{4}$ K is about 5 per
cent larger than a cold white dwarf. To correct from $10^{4}$ to
$14000$ K we have used the white dwarf cooling curves of
\citet{wood95}, which gives a total radial correction of 6.2 per
cent. This alternative model is also given in Table \ref{parameters},
and is quoted for all temperature-dependent parameters.

The expected flux from a blackbody $B_{\nu}(\lambda , T)$ in a
passband with transmission function $P(\lambda)$ is
\citep[e.g.][]{wood89a}
\begin{equation}
\label{eq:flux}
f= \frac{\int P(\lambda)B_{\nu}(\lambda , T)d\lambda/\lambda}{\int
  P(\lambda)d\lambda/\lambda}\cdot \frac{\pi R^{2}_{w}}{D^{2}},
\end{equation}
where $D$ is the distance to the star. By fitting a blackbody function
to the white dwarf flux in each passband, given in Table \ref{wd_times}, we
derive a white dwarf temperature of $T_{w}=13900 \pm 600$ K and
$D=51 \pm 17$ pc (assuming $R_{w}=0.0097 \pm 0.0031R_{\sun}$). We would
expect some systematic errors to be present in these estimates as we
have only used one eclipse and not an averaged lightcurve, so
flickering is likely to be a problem.

\begin{table*}
\begin{center}
\caption[]{System parameters of OU Vir. The secondary radius given is
  the volume radius of the secondary's Roche lobe \citep{eggleton83},
  as defined by equation \ref{eq:roche}. Parameters left blank in the
  right-hand column are independent of the model used.}
\begin{tabular}{lcc}
\hline
Parameter & Nauenberg (cold) & Nauenberg ($14000$ K)\\
\hline
Inclination $i$ & $79\fdg2 \pm 0\fdg7$ & \\
Mass ratio $q=M_{r}/M_{w}$ & $0.175 \pm 0.025$ & \\
White dwarf mass $M_{w}/M_{\sun}$ & $0.85 \pm 0.20$ & $0.89 \pm 0.20$\\
Secondary mass $M_{r}/M_{\sun}$ & $0.15 \pm 0.04$ & $0.16 \pm 0.04$\\
White dwarf radius $R_{w}/R_{\sun}$ & $0.0095 \pm 0.0030$ & $0.0097 \pm 0.0031$\\
Secondary radius $R_{r}/R_{\sun}$ & $0.177 \pm 0.024$ & $0.180 \pm 0.024$\\
Separation $a/R_{\sun}$ & $0.73 \pm 0.06$ & $0.74 \pm 0.05$\\
White dwarf radial velocity $K_{w}/\rmn{km\;s^{-1}}$ & $75 \pm 12$ & $76 \pm 12$\\
Secondary radial velocity $K_{r}/\rmn{km\;s^{-1}}$ & $426 \pm 6$ & $432 \pm 6$\\
Outer disc radius $R_{d}/a$ & $0.2315 \pm 0.0150$ & \\
Distance $D/\rm{pc}$ & $51 \pm 17$ & \\
White dwarf temperature $T_{w}/\rm{K}$ & $13900 \pm 600$ & \\
\hline
\end{tabular}
\label{parameters}
\end{center}
\end{table*}

\section{Conclusions}
\label{conclusions}

We have presented an analysis of 5 eclipses of OU Vir, some in
superoutburst and some in quiescence. The quiescent eclipses have been
used to make the first determination of the system parameters, given
in Table \ref{parameters}. Our main conclusions are as follows:
\begin{enumerate}
\item Eclipses of both the white dwarf and bright spot were observed
  during quiescence. The identification of the bright spot ingress and
  egress appears unambiguous.
\item By requiring the gas stream to pass directly through the light
  centre of the bright spot the mass ratio and orbital inclination
  were found to be $q=0.175 \pm 0.025$ and $i=79\fdg2 \pm 0\fdg7$.
\item Assuming that the central eclipsed object is circular, that its
  size accurately reflects that of the white dwarf and that it obeys
  the \citet{nauenberg72} approximation to the \citet{hamada61}
  mass-radius relationship, adjusted to $T=14000$ K, we find that the
  white dwarf radius is $R_{w}=0.0097 \pm 0.0031 R_{\sun}$ and its mass
  is $M_{w}=0.89 \pm 0.20 M_{\sun}$.
\item With the same assumptions, we find that the volume radius of the
  secondary star is $R_{r}=0.180 \pm 0.024 R_{\sun}$ and that its mass
  is $M_{r}=0.16 \pm 0.04 M_{\sun}$. The secondary star is therefore
  consistent with the empirical mass-radius relation for the
  main-sequence secondary stars in CVs of \citet{smith98a}.
\item A blackbody fit to the white dwarf flux gives a temperature
  $T_{w}=13900 \pm 600$ K and a distance $D=51 \pm 17$ pc with the same
  assumptions as above. These are purely formal errors from the
  least-squares fit using estimated errors of $\pm0.01$ mJy for each
  flux measurement. Given that we use data from only one eclipse, with
  a single measurement of the flux from each passband, the actual
  uncertainties are likely to be significantly larger.
\item The accretion disc radius of $R_{d}/a=0.2315 \pm 0.0150$ is
  similar in size to that of HT Cas, for which \citet{horne91b}
  derived $R_{d}/a=0.23 \pm 0.03$. This is small compared to many
  other dwarf novae (e.g.\ Z Cha, which has $R_{d}/a=0.334$;
  \citealt{wood89a}), but larger than the circularisation radius
  \citep[][ equation 13]{verbunt88} of $R_{circ}=0.1820a$. We note
  that this is an unusually small disc radius. It is especially
  surprising as this disc radius was determined from observations
  obtained only 20 days after the superoutburst was first reported
  \citep{kato03}.
\item The superhump period of OU Vir is $P_{sh}=0.078 \pm 0.002\rm$
  days \citep{vanmunster00}, which means OU Vir lies $5\sigma$ off the
  superhump period excess--mass ratio relation of \citet[][ equation
  8]{patterson98}, with the superhump period excess
  $\epsilon=(P_{sh}-P_{orb})/P_{orb}\sim0.073$. However, it does not
  lie on the superhump period excess--orbital period relation either,
  perhaps indicating that the current estimate of the superhump period
  $P_{sh}$ is inaccurate.
\end{enumerate}

\section*{acknowledgments}
We thank the anonymous referee for useful comments.
We are grateful to Tonny Vanmunster for providing his times of
mid-eclipse. WJF would like to thank Timothy Thoroughgood for helpful
discussions. WJF, MJS and CSB are supported by PPARC studentships.
CAW is employed on PPARC grant PPA/G/S/2000/00598. ULTRACAM is funded
by PPARC grant PPA/G/S/2002/00092.

\bibliographystyle{mn2e} \bibliography{abbrev,refs}

\begin{thebibliography}{}

\bibitem[\protect\citeauthoryear{Baptista, Catal\'an \& Costa}{Baptista
  et~al.}{2000}]{baptista00}
Baptista R.,  Catal\'an M.~S.,    Costa L.,  2000, MNRAS, 316, 529

\bibitem[\protect\citeauthoryear{Cook \& Warner}{Cook \& Warner}{1984}]{cook84}
Cook M.~C.,  Warner B.,  1984, MNRAS, 207, 705

\bibitem[\protect\citeauthoryear{Dhillon \& Marsh}{Dhillon \&
  Marsh}{2001}]{dhillon01b}
Dhillon V.~S.,  Marsh T.~R.,  2001, New~Ast.~Rev., 45, Issue 1-2, 91

\bibitem[\protect\citeauthoryear{Eggleton}{Eggleton}{1983}]{eggleton83}
Eggleton P.~P.,  1983, ApJ, 268, 368

\bibitem[\protect\citeauthoryear{Flannery}{Flannery}{1975}]{flannery75}
Flannery B.~P.,  1975, MNRAS, 170, 325

\bibitem[\protect\citeauthoryear{Fukugita, Ichikawa, Gunn, Doi, Shimasaku \&
  Schneider}{Fukugita et~al.}{1996}]{fukugita96}
Fukugita M.,  Ichikawa T.,  Gunn J.~E.,  Doi M.,  Shimasaku K.,    Schneider
  D.~P.,  1996, AJ, 111, 1748

\bibitem[\protect\citeauthoryear{Hamada \& Salpeter}{Hamada \&
  Salpeter}{1961}]{hamada61}
Hamada T.,  Salpeter E.~E.,  1961, ApJ, 134, 683

\bibitem[\protect\citeauthoryear{Horne, Wood \& Steining}{Horne
  et~al.}{1991}]{horne91b}
Horne K.,  Wood J.~H.,    Steining R.~F.,  1991, ApJ, 378, 271

\bibitem[\protect\citeauthoryear{Kato}{Kato}{2003}]{kato03}
Kato T.,  2003, vsnet-alert, No. 7733

\bibitem[\protect\citeauthoryear{King}{King}{1985}]{king85}
King D.~L.,  1985, RGO/La Palma Technical Note~31, Atmospheric Extinction at
  the Roque de los Muchachos Observatory, La Palma

\bibitem[\protect\citeauthoryear{{Koester} \& {Sch\"{o}nberner}}{{Koester} \&
  {Sch\"{o}nberner}}{1986}]{koester86}
{Koester} D.,  {Sch\"{o}nberner} D.,  1986, A\&A, 154, 125

\bibitem[\protect\citeauthoryear{Lubow \& Shu}{Lubow \& Shu}{1975}]{lubow75}
Lubow S.~H.,  Shu F.~H.,  1975, ApJ, 198, 383

\bibitem[\protect\citeauthoryear{Lubow \& Shu}{Lubow \& Shu}{1976}]{lubow76}
Lubow S.~H.,  Shu F.~H.,  1976, ApJ, 207, L53

\bibitem[\protect\citeauthoryear{Mason, Howell, Szkody, Harrison, Holtzman \&
  Hoard}{Mason et~al.}{2002}]{mason02}
Mason E.,  Howell S.~B.,  Szkody P.,  Harrison T.~E.,  Holtzman J.~A.,    Hoard
  D.~W.,  2002, A\&A, 396, 633

\bibitem[\protect\citeauthoryear{Nauenberg}{Nauenberg}{1972}]{nauenberg72}
Nauenberg M.,  1972, ApJ, 175, 417

\bibitem[\protect\citeauthoryear{Patterson}{Patterson}{1998}]{patterson98}
Patterson J.,  1998, PASP, 110, 1132

\bibitem[\protect\citeauthoryear{Rutten, Kuulkers, Vogt \& van Paradijs}{Rutten
  et~al.}{1992}]{rutten92c}
Rutten R. G.~M.,  Kuulkers E.,  Vogt N.,    van Paradijs J.,  1992, A\&A, 265,
  159

\bibitem[\protect\citeauthoryear{Smith \& Dhillon}{Smith \&
  Dhillon}{1998}]{smith98a}
Smith D.~A.,  Dhillon V.~S.,  1998, MNRAS, 301, 767

\bibitem[\protect\citeauthoryear{Smith, Tucker \& et al.}{Smith
  et~al.}{2002}]{smith02}
Smith J.~A.,  Tucker D.~L.,    et al. 2002, AJ, 123, 2121

\bibitem[\protect\citeauthoryear{Vanmunster, Velthuis \& McCormick}{Vanmunster
  et~al.}{2000}]{vanmunster00}
Vanmunster T.,  Velthuis F.,    McCormick J.,  2000, Inf.\ Bull.\ var.\ Stars,
  4955

\bibitem[\protect\citeauthoryear{Verbunt \& Rappaport}{Verbunt \&
  Rappaport}{1988}]{verbunt88}
Verbunt F.,  Rappaport S.,  1988, ApJ, 332, 193

\bibitem[\protect\citeauthoryear{Wood, Horne, Berriman \& Wade}{Wood
  et~al.}{1989}]{wood89a}
Wood J.~H.,  Horne K.,  Berriman G.,    Wade 1989, ApJ, 314, 974

\bibitem[\protect\citeauthoryear{Wood, Horne, Berriman, Wade, O'Donoghue \&
  Warner}{Wood et~al.}{1986}]{wood86b}
Wood J.~H.,  Horne K.,  Berriman G.,  Wade R.,  O'Donoghue D.,    Warner B.,
  1986, MNRAS, 219, 629

\bibitem[\protect\citeauthoryear{Wood, Irwin \& Pringle}{Wood
  et~al.}{1985}]{wood85}
Wood J.~H.,  Irwin M.~J.,    Pringle J.~E.,  1985, MNRAS, 214, 475

\bibitem[\protect\citeauthoryear{Wood}{Wood}{1995}]{wood95}
Wood M.~A.,  1995, in Koester D.,  Werner K.,  eds, European Workshop on White
  Dwarfs, Lecture Notes in Physics, Volume 443, p.~41

\end{thebibliography}

\setcounter{table}{4}
\begin{table*}
\begin{center}
\caption[]{System parameters of OU Vir. The secondary radius given is
  the volume radius of the secondary's Roche lobe (Eggleton 1983),
  as defined by equation~(6). Parameters left blank in the
  right-hand column are independent of the model used.}
\begin{tabular}{lcc}
\hline
Parameter & Nauenberg (cold) & Nauenberg ($21\,700$~K)\\
\hline
Inclination $i$ & $79\fdg2 \pm 0\fdg7$ & \\
Mass ratio $q=M_{r}/M_{w}$ & $0.175 \pm 0.025$ & \\
White dwarf mass $M_{w}/M_{\sun}$ & $0.85 \pm 0.20$ & $0.90 \pm 0.19$\\
Secondary mass $M_{r}/M_{\sun}$ & $0.15 \pm 0.04$ & $0.16 \pm 0.04$\\
White dwarf radius $R_{w}/R_{\sun}$ & $0.0095 \pm 0.0030$ & $0.0097 \pm 0.0031$\\
Secondary radius $R_{r}/R_{\sun}$ & $0.177 \pm 0.024$ & $0.181 \pm 0.024$\\
Separation $a/R_{\sun}$ & $0.73 \pm 0.06$ & $0.75 \pm 0.05$\\
White dwarf radial velocity $K_{w}/\rmn{km\;s^{-1}}$ & $75 \pm 12$ & $76 \pm 12$\\
Secondary radial velocity $K_{r}/\rmn{km\;s^{-1}}$ & $426 \pm 6$ & $434 \pm 6$\\
Outer disc radius $R_{d}/a$ & $0.2315 \pm 0.0150$ & \\
Distance $D/\rm{pc}$ & $650 \pm 210$ & \\
White dwarf temperature $T_{w}/\rm{K}$ & $21700 \pm 1200$ & \\
\hline
\end{tabular}
\end{center}
\end{table*}

\section*{Erratum}
The paper `ULTRACAM photometry of the eclipsing cataclysmic variable
OU Vir' was published in Mon.\ Not.\ R.\ Astron.\ Soc., {\bf 347},
1173-1179 (2004). We have since discovered two errors. First, the
observations of 2002 May 16 shown in the upper two panels of Figure 1
have incorrect fluxes due to the adoption of an incorrect CCD gain
value during commissioning. The {\em y}-axis scale (and all vertical
offsets) should be divided by a factor of 4.1 to get the correct
fluxes. This places the system at its quiescent brightness, comparable
to our subsequent observations shown in the lower four panels of
Figure 1. We have stated in sections 3 and 7 that the system was in
decline from superoutburst in 2003 May, which the flux no longer
supports (although there is a feature that

\newpage 
\section*{}
resembles a superhump at phase $\sim -0.45$ similar to that observed
about 18 days after the start of the 2003 superoutburst).
Second, the blackbody fit discussed at the end of section 6 was
calculated incorrectly due to a programming error. The correct fit
yields a white dwarf temperature $T_{\rmn{w}}=21700\pm1200$~K and a
distance to the star $D=650\pm210$~pc. The new temperature estimate
results in a radial correction to the white dwarf of 8.0~per~cent,
which has a negligible knock-on effect on the system parameters listed
in Table~5, the abstract and the conclusions (section~7). The new,
corrected Table~5 is presented above. All other results are
unaffected. We thank Dr.\ Boris G{\"{a}}nsicke for bringing the latter
problem to our attention.

\end{document}